\documentclass[iop,twocolumn]{aastex62}

\usepackage{graphicx}
\usepackage{amssymb}
\usepackage{amsmath}
\usepackage{natbib}
\usepackage{longtable}
\usepackage{footnote}
\usepackage{helvet}
\usepackage{color}
\usepackage{afterpage}
\usepackage{float}
\usepackage{url}

\newcommand{\ayushiemail}{ayushi.singh@mail.utoronto.ca}

\newcommand{\calT}{{\cal T}}
\newcommand{\calW}{{\cal W}}
\newcommand{\calM}{{\cal M}}
\newcommand{\calR}{{\cal R}}
\newcommand{\rvec}{{\mathbf r}}
\newcommand{\wvec}{{\mathbf w}}
\newcommand{\gvec}{{\mathbf g}}
\newcommand{\Sigmacl}{\Sigma_{\rm cl}}
\newcommand{\Sigmaenv}{\Sigma_{\rm env}}
\newcommand{\Sigmabg}{\Sigma_{\rm bg}}

\newcommand{\psitwod}{{\psi_{\rm cl,2D}}}
\newcommand{\calWcl}{{\cal W}_{\rm cl}}
\newcommand{\calWtwod}{{\cal W}_{\rm cl,2D}}
\newcommand{\calTtwod}{{\cal T}_{2D}}
\newcommand{\sigmaclz}{\sigma_{{\rm cl},z}}
\newcommand{\sigmath}{\sigma_{\rm th}}
\newcommand{\sigmaNT}{\sigma_{{\rm NT},z}}
\newcommand{\rhothresh}{\rho_{\rm edge}}

\newcommand{\Rcl}{R_{\rm cl}} 
\newcommand{\Mcl}{M_{\rm cl}} 
\newcommand{\vCMz}{v_{{\rm CM},z}} 
\newcommand{\alphaBM}{\alpha_{\rm BM92}}
\newcommand{\alphacl}{\alpha_{\rm cl}}
\newcommand{\alphatwod}{\alpha_{\rm 2D}}

\begin{document}

\title{VIRIAL RATIO: DIRECT EVALUATION FROM MOLECULAR CLOUD DATA AND THE CHALLENGES OF IMPROVING ACCURACY}

\correspondingauthor{Ayushi Singh}
\email{\ayushiemail}

\shorttitle{Virial ratios of clouds}
\shortauthors{Singh et al.}

\author{Ayushi Singh }
\affil{Department of Astronomy and Astrophysics, University of Toronto, 50 St. George Street, Toronto, Ontario, M5S 3H4, Canada}
\author{Christopher D. Matzner}
\affil{Department of Astronomy and Astrophysics, University of Toronto, 50 St. George Street, Toronto, Ontario, M5S 3H4, Canada}
\author{Peter H. Jumper}
\affil{10 Jody Lane, Forestdale, MA 02644, USA}

 \accepted{in {\em ApJ}, April 2019}

\begin{abstract}
{The virial ratio between  kinetic and gravitational terms provides key insight into the balance of forces that confine a molecular cloud, but the clumpy and filamentary structures of resolved clouds make it difficult to evaluate this ratio in a consistent way.  For clouds with resolved maps of column density as well as a line tracer, we demonstrate that the gravitational energy can be estimated directly from observations in a manner similar to the kinetic energy.  This offers improved diagnostic power and consistency.   Disentangling a cloud from foreground and background materials is a persistent challenge, for which we introduce a strategy based on Abel's transform.  We provide proofs of principle using simulated clouds.  }
\end{abstract}


\section{Introduction}

Interpretations of molecular cloud observations often rely on the virial ratio 
\begin{equation} \label{alpha_def} 
\alphacl \equiv {2\calT\over |\calWcl|}
\end{equation} 
between kinetic ($\calT$) and self-gravitational ($\calWcl$) energies.  These are the most easily observable terms in the virial theorem, which relates radial accelerations to four energetic terms: 
\begin{equation}\label{eq:Virial}
\frac12 \ddot{I}= 2(\calT-\calT_0)+ \calW  +\calM+ \calR
\end{equation}
(\citealt{Clausius1870, ChandrasekharFermi53}; here in Lagrangian form \citealt{McKee1992}). Here $I=\int r^2\, dm$ is the trace of the cloud's  moment-of-inertia tensor (relative to its center of mass at $r=0$); $\calT$ is the total (bulk plus thermal) kinetic energy in the center-of-mass frame; $\calT_0 = \frac12 \int P\,{\mathbf r}\cdot {\mathbf dS}$ represents confinement by external pressure on the cloud boundary; $\calW = \int {\mathbf r}\cdot {\mathbf g}\, dM$ is the (negative) gravitational energy (of which $\calWcl$ is a part: see \S\,\ref{S:GravEstimation}), and $\calM$ and $\calR$ are the terms due to magnetic and (usually  negligible) radiation forces, respectively.  

Equilibrium interstellar structures, if strongly self-gravitating and supported by thermal pressure, have virial ratios of order two or somewhat higher; for instance, $\alphacl = 2.1$ in the critical Bonnor--Ebert sphere.   Insofar as observed clouds are not far from equilibrium ($\ddot{I} < |\calW|$) and radiation pressure is negligible, values well below two indicate strong magnetic support (high $\calM/|\calW|$), while values well above two indicate strong confinement by external pressure (high $2\calT_0/|\calW|$), as might result from the weight of an atomic envelope or the ram pressure of exterior flow. Likewise, a region with $\alphacl>2$ is energetically unbound, in the sense that the sum of self-gravitational, magnetic, and kinetic energies must be positive in the center-of-mass frame. 
In practice, the virial ratio $\alphacl$ is usually estimated by means of the virial parameter introduced by \citet{Bertoldi1992},
\begin{equation}\label{eq:alpha}
\alpha_{\rm BM92} \equiv \frac{5\sigmaclz^2 \Rcl}{G \Mcl},
\end{equation} 
This is especially convenient because it only requires one to evaluate the cloud's mass $\Mcl$, effective radius $\Rcl$, and line-of-sight velocity dispersion $\sigmaclz$. Its relation to $\alphacl$ is derived from two observations. First, if the velocity dispersion along the line of sight is representative of the other two directions, then  $2\calT\simeq 3\Mcl\sigmaclz^2$. Second, if one defines the parameter $a$ by  $\calWcl = -\frac35 a G\Mcl^2/\Rcl$, concentrated spheroidal structures with a wide range of axis ratios satisfy $1\leq a \leq2$  \citep{Bertoldi1992}. Combining these, $\alphacl \simeq \alphaBM/a$.   

As a practical example of its use, consider the decreasing trend of $\alphaBM$ with decreasing mass and increasing column density in star-forming molecular regions.  The virial parameter frequently exceeds two in the regions identified as giant molecular clouds \citep{Heyer2009}, whereas it drops to well below unity in the dense, filamentary regions representative of massive star and star cluster formation (e.g.,  \citealt{Bertoldi1992,Kauffmann2013,2018MNRAS.477.2220T}), notwithstanding the existence of apparently pressure-confined substructures
\citep[e.g.][]{Bertoldi1992,Chen18droplets}. 
This trend has been interpreted in terms of pressure confinement on GMC scales and incipient or ongoing collapse on star cluster formation scales \citep{Field2011, Kauffmann2013, Matzner2015, 2018MNRAS.473.4975T} in which turbulent motions are fed by gravitational compression \citep{2009ApJ...707.1023V, 2011MNRAS.411...65B, 2011ApJ...738..101G, 2012ApJ...750L..31R}. These conclusions are consistent with modern simulations of well-resolved galactic disks with supernova feedback.  

Making precise comparisons requires estimating $\alphacl$ across a range of environments and observations.  However, this becomes difficult when high-resolution data reveal that the cloud of interest is composed of clumpy and filamentary structures;  this makes the assignment of a cloud boundary somewhat arbitrary and the value of $\Rcl$ somewhat uncertain. Various authors use the size of a model fit to the data, or a width at half maximum, or the square root of the enclosed area to evaluate $\Rcl$. Studies are not always consistent about the other terms: $\sigmaclz$ is often taken to be a typical value of the local line width, even in the presence of significant velocity gradients, and approaches differ on whether all the column projected within the cloud boundary should be counted toward $\Mcl$, or whether $\Mcl$ should be taken from a fit to the data (and if so, which profile to fit).  Finally, if $\alphacl$ is to be inferred from $\alphaBM$, then $a$ is sometimes estimated from the apparent density profile but otherwise  assumed (explicitly or tacitly) to take a certain value.  

Our goal, therefore, is to offer a relatively consistent strategy for estimating $\calWcl$ and $\cal T$, and hence $\alphacl$, directly from the available observational data given a user-defined cloud boundary.   The most novel aspect of our proposal is a way to estimate $\calWcl$  directly from cloud column density maps (\S~\ref{S:GravEstimation}) with essentially the same fidelity as the evaluation of $\calT$. We also recommend an evaluation of $\cal T$ (albeit not a novel one) in \S~\ref{SS:KineticEstimation}.  Finally, we propose a reasonably model-independent method (\S~\ref{SS:Abel-theory}), based on the Abel transform, to remove foreground and background material associated with the region of interest.  Our proposed approach eliminates some ambiguity, while highlighting the fact that the viewing angle dependence, and the removal of foreground and background matter, remain limiting factors in the evaluation of $\cal T$, $\cal W$, and $\alphacl$. 

\section{Direct estimation of virial terms}

We now turn to how to estimate terms in the virial theorem directly from observations. First, we define our terms. 

We consider the cloud volume $V_{\rm cl}$ (coordinates $\rvec = [x,y,z]$, with $z$ along our line of sight), defined in three-dimensional space by its density distribution $\rho(\rvec)$ and, as observed, by its projected column density $\Sigmacl(x,y) = \int_{V_{\rm cl}} \rho\, dz$. When projected onto the plane of the sky, the cloud is enclosed by the area $A_{\rm cl}$.  The cloud mass is $\Mcl = \int_{V_{\rm cl}}\rho\, d^3\,\rvec =  \int_{A_{\rm cl}} \Sigmacl\,dx\,dy$. The mass-averaged radial velocity along each line of sight is $v_z(x,y)$, so the line of sight velocity of the cloud's center of mass is  $\vCMz = \Mcl^{-1} \int v_z\Sigmacl\,dx\,dy$.  The thermal velocity dispersion  $\sigmath(x,y)$ represents all particles along each line of sight (not just the line-emitting particles) and is defined so that $\frac32 \Sigmacl \sigmath^2$ is the thermal kinetic energy per unit area.  The nonthermal velocity dispersion $\sigmaNT(x,y)$ indicates variations in the bulk radial velocity within each resolution element, including velocity differences along the sight line.

\subsection{Gravitational energy}
\label{S:GravEstimation}
The term $\calW$ is composed of two parts: the cloud's gravitational self-energy, $\calWcl = \int_V \gvec_{\rm cl}\cdot \rvec \,d^3\,\rvec$ where $\gvec_{\rm cl}$ is the gravity of cloud matter alone, and a term due entirely to tidal gravity from the cloud's environment, ${\cal W}_{\rm ext} = \int_V (\gvec-\gvec_{\rm cl})\cdot \rvec \,d^3\,\rvec$. 
 
Following tradition, we define $\alphacl$ in terms of $\calWcl$ alone; this is also the reason we call it $\alphacl$ rather than simply $\alpha$.  We note, however, that $\calW_{\rm ext}$ can be non-negligible for clouds whose densities are not far above the Roche limit and that it is usually positive. 

While uncertainty about the distribution of matter along the line of sight  prevents one from obtaining $\calWcl$ directly from $\Sigmacl$, we can compute a similar quantity: $\calWtwod$,  which is $2/\pi$ times the value $\calWcl$ would have if the cloud were collapsed to a thin sheet (at the same distance) in the plane of the sky.  (We use the subscript `2D' to indicate a quantity derived from projection along one line of sight.)  Mathematically, 
\begin{equation} \calWtwod = \frac1\pi \int \Sigmacl \psitwod\,dx\,dy \end{equation}
where $\psitwod$ is the corresponding potential. 

\begin{figure}[t]
\centering
$\begin{array}{c}
    \vspace{-2cm}
    \hspace{-1cm}
    \includegraphics[scale=0.25]{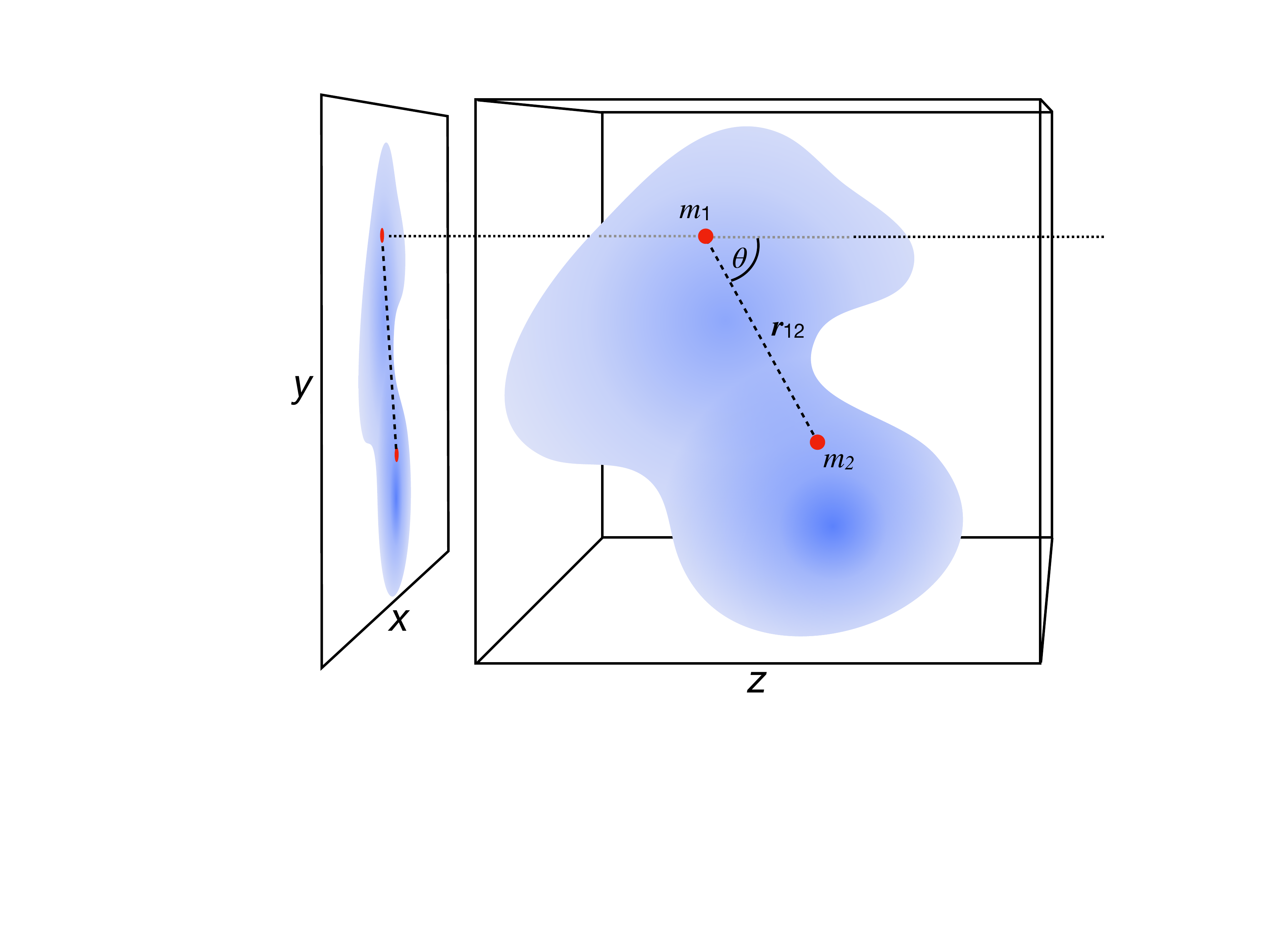} \\
\end{array}$
\caption{To estimate the gravitational energy $\cal W$ of a three-dimensional cloud from its observed column density, we calculate the energy it would have if along the line of sight into a sheet (dotted line).  Averaged over orientations, the energy of the sheet is exactly $\pi/2$ times greater than the cloud energy (\S~\ref{S:GravEstimation}).  This result holds for each pair of masses comprising the cloud, illustrated by $m_1$ and $m_2$ in this schematic.  }
\label{fig:GravSchematic}
\end{figure}

A main result of our paper is the definite relation between $\calWcl$ and $\calWtwod$: 
\begin{equation}\label{eq:2d-to-3d-ratio}
 \calWcl=
 \left<  \calWtwod \right>  
\end{equation}
where $\left<\cdots\right>$ means an average over orientations.  The proof is simple: the gravitational energy of a three-dimensional mass configuration equals the sum of pairwise energies for all pairs of particles within it and the same is true of the flattened configuration.   Two such particles of masses $m_1$ and $m_2$, separated by distance $r_{12}$, have pairwise energy $-Gm_1m_2/r_{12}$.   If we project them  along $\mathbf{\hat{z}}$ on to a sheet, such that $\mathbf{\hat{z}} \cdot \hat \rvec_{12} =\cos \theta = \mu$, as in Figure \ref{fig:GravSchematic}, the separation within this sheet is $(1-\mu^2)^{1/2} r_{12}$ and the gravitational energy within the sheet is larger than in the cloud by the factor $(1-\mu^2)^{-1/2}$.   Averaging over angles gives the factor $\frac12\int_{-1}^1 (1-\mu^2)^{-1/2}\,d\mu = \pi/2$ which cancels the prefactor $2/\pi$ we introduced in defining $\calWtwod$.
A detailed derivation is presented in the Appendix.

As a corollary, any spherically symmetric cloud satisfies $\calWtwod=\calWcl$ identically.  Checking both spherical and simulated three-dimensional density distributions, we find perfect agreement within discretization errors.  

By relating the observable  $\calWtwod$ to the desired quantity $\calWcl$, equation (\ref{eq:2d-to-3d-ratio}) sidesteps the need to adopt values for $\Rcl$ and $a$ in estimates based on $\alphacl$ (although one must still specify a two-dimensional cloud boundary).  Furthermore, $\psitwod$ and $\calWtwod$ are easy to compute from $\Sigmacl$ using convolution with the appropriate Green's function: \[ \psitwod = -{\frac{G}{\sqrt{x^2+y^2}}}\circledast \Sigmacl.\] 

Although this is a step toward directly evaluating  $2\calT/|\calW|$, several issues remain.  First, one must estimate $\calT$ from an additional data set.  Second, one must somehow identify a projected cloud boundary and separate $\Sigmacl$ from the rest of $\Sigma$, which usually includes foreground and background emission (partly in a physically associated cloud envelope). We comment on these issues below.

\subsection{Kinetic energy} \label{SS:KineticEstimation} 
When evaluating $\calT$, one is limited by a lack of knowledge about motions in the plane of the sky, as only the thermal kinetic energy is guaranteed to be isotropic.  However, defining $\calTtwod$ to be triple the kinetic energy in line-of-sight motions,
\begin{equation} \label{eq:KineticEnergy}
\calTtwod = \frac32 \int \left[\sigmath^2 + \sigmaNT^2 + (v_z-\vCMz)^2\right] \Sigmacl \,dx \,dy, 
\end{equation}
ensures that 
\begin{equation} \label{eq:avgKE} 
\calT = \left<\calTtwod\right>,
\end{equation}
and in this sense $\calTtwod$ is an optimal estimate for $\calT$.   Accordingly, $\sigmaclz$ is best defined via $\frac32 \Mcl \sigmaclz^2=\calTtwod$.

Comparing equations (\ref{eq:2d-to-3d-ratio}) and (\ref{eq:avgKE}), we see that observational estimates of $\calWcl$ and $\calT$ are on exactly the same footing, in the sense that we have an observational estimate for each term whose average over orientations or viewing angles yields the correct value. 

\subsection{Estimation of the virial ratio} \label{SS:VirialRatio}
These relations motivate the approximation
\begin{equation} \label{eq:VirialRatioEstimate} 
\alphacl  ~ \simeq~ \alphatwod \equiv {2\calTtwod\over |\calWtwod|} 
\end{equation} 
in which the numerator and denominator are separately correct when averaged over orientations. 
While $\left<2\calT/ \calWcl\right>$ is not equivalent to $2\left<\calT\right>/\left<\calWcl\right>$, equation $\alpha$ will be reasonably represented by $\alphatwod$ so long as the dependence on viewing angle is not too strong.  Furthermore, we anticipate that its error can be characterized in terms of properties of the observed maps, given prior expectations regarding the distribution of cloud properties.   We explore these errors using numerical simulations in  \S~\ref{S:Calibration_from_Sims}.  

\subsection{Application to an ensemble of clouds} \label{SS:CloudEnsembleAverage}
Insofar as each cloud is viewed at a random orientation, the net kinetic energy $\calT_{\rm tot} = \sum_i \calT_i$ and the net gravitational energy $\calW_{\rm tot} = \sum_i \calW_{{\rm cl,}i}$ of an ensemble of clouds (labeled $i$) are approximated by $\sum_i \calT_{2D,i}$ and $\sum_i \calW_{{\rm cl,2D},i}$ respectively, with increasing accuracy as the cloud population grows.  
For this reason, the net virial ratio 
$\alpha_{\rm tot} = 2\calT_{\rm tot}/|\calW_{\rm tot}|$ and the total (kinetic and gravitational) energy $\calT_{\rm tot} + \calW_{\rm tot}$ can be determined without much of the variance caused by projection effects.     

Nevertheless, a few caveats still apply to ensemble-based calculations. Relatively similar-mass clouds should be considered; otherwise, a small number of the most massive objects will tend to dominate the total energy.   Selection effects may be important, as objects are more easily identified if projected along a long axis.   Finally, there could exist some organizing structure, like the  magnetic field, Galactic disk, or spiral arms \citep[e.g.,][]{2006ApJ...638..191K}, that prevents an average over the population from also being an average over orientations.  The last effect should be detectable in correlations between cloud properties and sky position.  

\section{Disentangling column densities} \label{S:DisentanglingSigma}

Another complication of having only projected data is that no outline on the sky can specify what three-dimensional boundary encloses the cloud of interest.  The line-of-sight ambiguity is less serious if there is evidence of a single density maximum.   Regardless, the two-dimensional boundary is guaranteed to include foreground and background material, some of which is physically associated to, and contiguous with, the cloud of interest.  

We focus here on how to estimate the cloud column directly from $\Sigma$ and a chosen cloud outline -- which of course should be covered by the molecular emission used to derive velocities and temperatures. (We set aside for now the fact that line emission, by virtue of its density dependence, may allow one to sidestep some of the difficulties associated with projection,  while introducing uncertainties involving tracer abundance, line excitation, and radiative transfer.) 

We therefore decompose $\Sigma$ into three components: the cloud itself, a physically associated cloud ``envelope", and unrelated foreground or background matter: 
\[ \Sigma = \Sigmacl+\Sigma_{\rm env} + \Sigma_{\rm bg}.  \] 
While this decomposition is not unique, several points guide an estimate for $\Sigmacl$ if we assume that our projected cloud boundary contains a peak in $\Sigma$: 
\begin{itemize}
\item[$i-$]{$\Sigmacl$ continuously approaches zero toward the cloud boundary, because lines of sight at the cloud edge are tangent to a curved surface; and $\Sigmacl$ is zero outside the boundary. } 
\item[$ii-$]{$\Sigmacl + \Sigma_{\rm env}$ is a continuous function of position that increases toward a local maximum within the cloud boundary, and correlates with molecular line emission from the cloud and envelope.}
\item[$iii-$]{$\Sigma_{\rm bg}$ does not correlate with the other components and their molecular line emission.}
\item[$iv-$]{Each component of $\Sigma$ is nonnegative. }
\end{itemize}

These considerations provide some guidance regarding how to estimate $\Sigmacl$ given only $\Sigma$ and a two-dimensional cloud boundary.  The very simplest is to ascribe the total column to the cloud and adopt $\Sigmacl=\Sigma$ within the boundary.   This is guaranteed to be an overestimate, however, if $\Sigma$ is nonzero on the boundary.   The next level of sophistication is to estimate $\Sigmacl$ by subtracting a smooth function (representing $\Sigma_{\rm env}+\Sigma_{\rm bg}$) that matches $\Sigma$ at the  boundary.   While this guarantees $\Sigmacl=0$ at the boundary, it tends to underestimate the central values of $\Sigmacl$ because $\Sigma_{\rm env}$ tends to be greater at the edge than in the boundary, i.e., limb brightened.     A  further refinement would be to account for this effect by modeling the contribution of the envelope within the cloud boundary, perhaps using an idealized hydrostatic model such as a Bonnor--Ebert sphere.   

Finally, one can hope to recover the density distribution of the cloud and its envelope by analyzing the data in a relatively model-independent way, and use this reconstruction to isolate $\Sigmacl$.    We propose one such reconstruction based on the Abel transform. 

\subsection{Abel transform reconstruction} \label{SS:Abel-theory} 

Abel's transform \citep{Abel1826} allows one to reconstruct any axisymmetric density distribution $\tilde\rho(\tilde r)$ from its projection  along a line orthogonal to the symmetry axis, $\tilde\Sigma(\tilde R)$: 
\begin{equation}\label{eq:abelbackward}
   \tilde{\rho} (\tilde{ r} ) = - \frac { 1 } { \pi } \int _ { \tilde{r} } ^ { \infty } \frac {d \tilde{\Sigma}/d\tilde{R}}   { \sqrt { \tilde{R} ^ { 2 } - \tilde{r} ^ { 2 } } } d \tilde{R},
\end{equation}
and vice versa,
\begin{equation}\label{eq:abelforward}
   \tilde{\Sigma} ( \tilde{R} ) =  2 \int _ { \tilde{R} } ^ { \infty } \frac { \tilde{\rho}({\tilde{r}})\tilde{r} } { \sqrt { \tilde{r} ^ { 2 } - \tilde{R} ^ { 2 } } } d \tilde{r}. 
\end{equation}
We use tildes for the axisymmetric problem to which Abel's transform applies, to distinguish it from the fully three-dimensional problem. 

A couple points motivate an attempt to adapt Abel's transform to the more general problem of reconstructing $\Sigmacl(x,y)$ from $\Sigma(x,y)$ and a cloud boundary contour.  
First, it is straightforward to reconstruct an axisymmetric cloud from column density data using the above equations. (Obtain $\tilde\rho(\tilde r)$ from equation \ref{eq:abelbackward}; identify the cloud radius $\Rcl$ from the corresponding boundary on the map;  set $\tilde\rho(\tilde r)=0$ for $\tilde r > \Rcl$ to obtain $\tilde \rho_{cl}(\tilde r)$; then use equation \ref{eq:abelforward} to determine $\tilde\Sigma_{\rm cl}$.)
Second, the reconstructed $\Sigmacl$ is accurate even if the line of sight is inclined, rather than normal to the symmetry axis.   Moreover, this reconstruction has the desired properties  $i$-$iv$ listed above.   

Our strategy will be to distill our two-dimensional data into a one-dimensional function $\tilde\Sigma(\tilde R)$, use the transform to construct the function $\tilde\Sigma_{\rm cl}/\tilde\Sigma_{\rm env}$ as a function of $\tilde \Sigma$ relative to a chosen threshold value, and then apply this function to the two-dimensional data to estimate $\Sigmacl/\Sigmaenv$.  We work under the assumption that the cloud boundary contains a maximum of $\Sigma$. 

As a first step, we subtract from $\Sigma$ a smooth function that does not correlate with the cloud, such as a low-order polynomial, to represent $\Sigmabg$.  The result is an estimate for $\Sigmacl + \Sigmaenv$, although we continue to call it $\Sigma$ for simplicity.  We then restrict attention to a region larger than the cloud but associated with the cloud peak, in the sense that 
$\nabla \Sigma$ points inward toward the cloud. 

To make our one-dimensional structure we flatten our maps of $\Sigma$ and $\nabla \Sigma$ into one-dimensional lists, then sort them by $\Sigma$ to create $\tilde{\Sigma}$ and $|\nabla \ln \tilde\Sigma|$. (Here, a tilde indicates one-dimensional structure.)  The list  $|\nabla \ln\tilde\Sigma|$ tends to correlate with $\tilde\Sigma$,  but with significant scatter that arises because nonadjacent data is brought together by the sorting process.   However, we find that smoothing this sorted list using a Gaussian window with a width of a few percent of the list length suffices to create an effectively continuous function: $|\nabla \ln\tilde\Sigma|(\tilde \Sigma)$.   (We work with the logarithmic gradient rather than the ordinary gradient because we wish the smoothing to represent a geometric mean rather than an arithmetic one.)  From this we create the effective radius by numerical integration: 
\begin{equation}\label{eq:effr}
    \tilde{R} = - \int_{0}^{\tilde{\Sigma}} \frac{d \ln \tilde{\Sigma}}{ |\nabla \ln \tilde{\Sigma}| } + C, 
\end{equation}
setting $C$ so that $\tilde R=0$ at the maximum of $\tilde\Sigma$.  The negative sign arises from the condition that $\nabla \Sigma$ points toward the cloud peak in the region of interest. 

We can now obtain  $\tilde{\rho}(\tilde R)$ from equation (\ref{eq:abelforward}), truncate it at the 
radius for which $\tilde\Sigma(\tilde R)= \Sigma_{\rm edge}$ representing the average value on the cloud boundary), 
and then use equation (\ref{eq:abelbackward}) to derive  $\tilde\Sigma_{\rm cl}$. 
Our ultimate goal is to determine $\tilde\Sigma_{\rm cl}/\tilde \Sigma$ as a function of $\tilde \Sigma/\Sigma_{\rm edge}$, and to apply this to function to the two-dimensional data.   In this application we estimate $\Sigma_{\rm edge}(x,y)$ as an interpolation of values from the cloud boundary. 

The result should be a method of background and envelope subtraction that is relatively model-independent, satisfies our requirements for a valid reconstruction, and is quite accurate for the case of an axisymmetric cloud observed from the side.   We test these expectations in \S~\ref{SS:ProjectionEffectsInSims}. 

\section{Calibration using computed clouds}
\label{S:Calibration_from_Sims}

We  wish to verify and calibrate equation (\ref{eq:VirialRatioEstimate}) and the techniques for disentangling $\Sigmacl$ from $\Sigma$ discussed in \S\,\ref{S:DisentanglingSigma}.  For a proof of principle we consider here only a few low-resolution numerical  simulations of self-gravitational collapse, saving more extensive calculations and comparison to observations to future works. We perform nonmagnetized, effectively isothermal (adiabatic index 1.0001) simulations within the Flash code \citep{Fryxell2000} at fixed {64}$^3$ resolution, with diode outflow boundary conditions, starting with a uniform spherical cloud, 25 cells in radius, within an atmosphere that is under-dense by a factor of 100 and initially in pressure equilibrium.   The initial Jeans length is 1.78 initial cloud radii, which suffices to guarantee collapse of the cloud core despite a rarefaction wave from the edge.  An initial velocity field of Mach number {0.29} in random cell-to-cell motions is imposed. This decays rapidly; but the velocity is also stirred with solenoidal accelerations on wavelengths of order the box length (and with an  autocorrelation time of 4.2 initial freefall times), at a rate sufficient to cause the collapsing cloud to become flattened and filamentary during its collapse.   Although our simulation allows for the creation of sink particles in regions whose density violates the Jeans criterion \citep{1997ApJ...489L.179T}, we are careful to only consider outputs prior to sink particle creation in order to avoid having to account for the particles' gravity.   The isothermal sound speed is $\sigmath$; dimensional scales of length, mass, and time are not relevant in our analysis.  

From each simulation we take output at { evenly spaced times after the peak density exceeds the threshold density discussed below, but before the first sink particle forms}, and for each of these times we generate three separate projections by choosing which axis is along the line of sight.   

Within our simulation we consider the cloud  to be the three-dimensional volume above a threshold density $\rhothresh$, which we set an order of magnitude above the mean simulation density to ensure the cloud is strongly condensed. (This definition allows for the cloud to break into separate subvolumes, although this does not in fact happen.) 

From this volume we compute projected maps.  For projection along $z$ these include $\Sigma(x,y) = \int \rho\,dz$, the total column density, and $\Sigmacl(x,y) = \int \rho\,\Theta_{\rm cl, 3d}\,dz$.  (The integral is performed as a sum over cells, and the mask function $\Theta_{\rm cl, 3d}= \Theta(\rho-\rhothresh)$ is unity within the cloud volume, zero elsewhere.)   Then 
$\Mcl {\mathbf v}_{\rm cm}  = \int \rho {\mathbf v}  \Theta_{\rm cl, 3d} \,d^3{\mathbf r}$ and 
\[ 2\calT  =3 \Mcl \sigmath^2 +  \int |{\mathbf v} - {\mathbf v}_{\rm cm}|^2 \rho  \,\Theta_{\rm cl, 3d}\, d^3{\mathbf r}. \]
Similarly the mass-averaged line-of-sight cloud velocity is $\bar v_z(x,y) = \int v_z \rho\Theta_{\rm cl, 3d}\,dz/\Sigmacl$, and the nonthermal velocity dispersion along the line of sight $\sigmaNT(x,y)$ is given by 
\[\sigmaNT^2 = \Sigmacl(x,y)^{-1} \int (v_z-\bar v_z)^2 \rho \, \Theta_{\rm cl, 3d} \,dz.\] 

\subsection{Inference of of virial terms and the virial ratio} \label{SS:VirialRatioInSims} 

\begin{figure}[t]
\centering
$\begin{array}{c}
    \vspace{0.2cm}
    \hspace{-0.2cm}
    \includegraphics[scale=0.35]{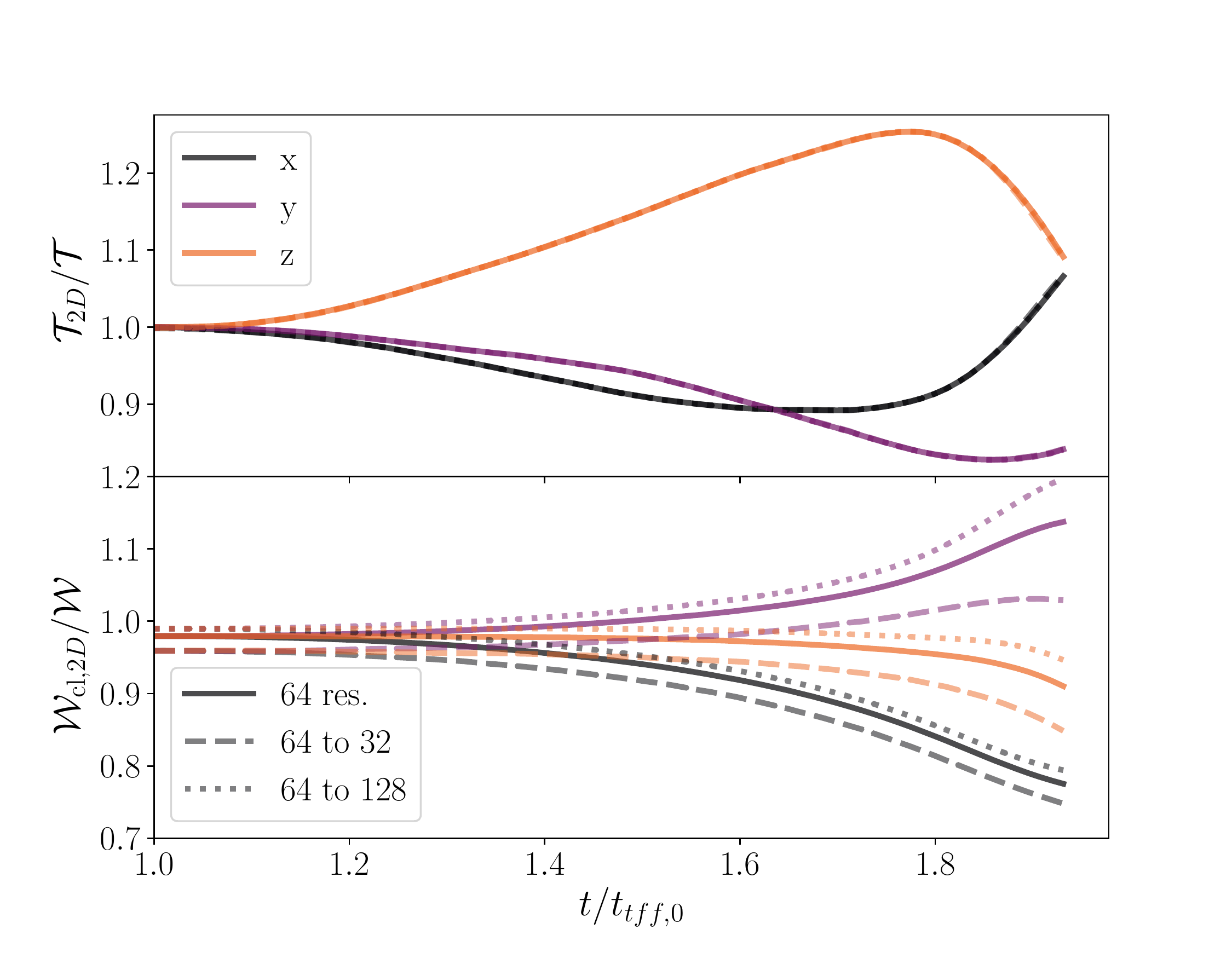} \\
\end{array}$
\caption{Apparent gravitational energy $\calWtwod$ and apparent kinetic energy $\calTtwod$ are compared to the ideal values ($\calWcl$ and $\calT$, respectively) for three orthogonal projections of the same simulation.  To uncover the effect of finite resolution on these quantities, we make the same comparison after degrading the native 64$^3$ resolution to 32$^3$ (dashed lines) or enhancing it via interpolation to $128^3$ (dotted lines). }
\label{fig:energy_timescale}
\end{figure}

\begin{figure}[ht]
\centering
$\begin{array}{c}
    \vspace{0.2cm}
    \hspace{-0.3cm}
    \includegraphics[scale=0.39]{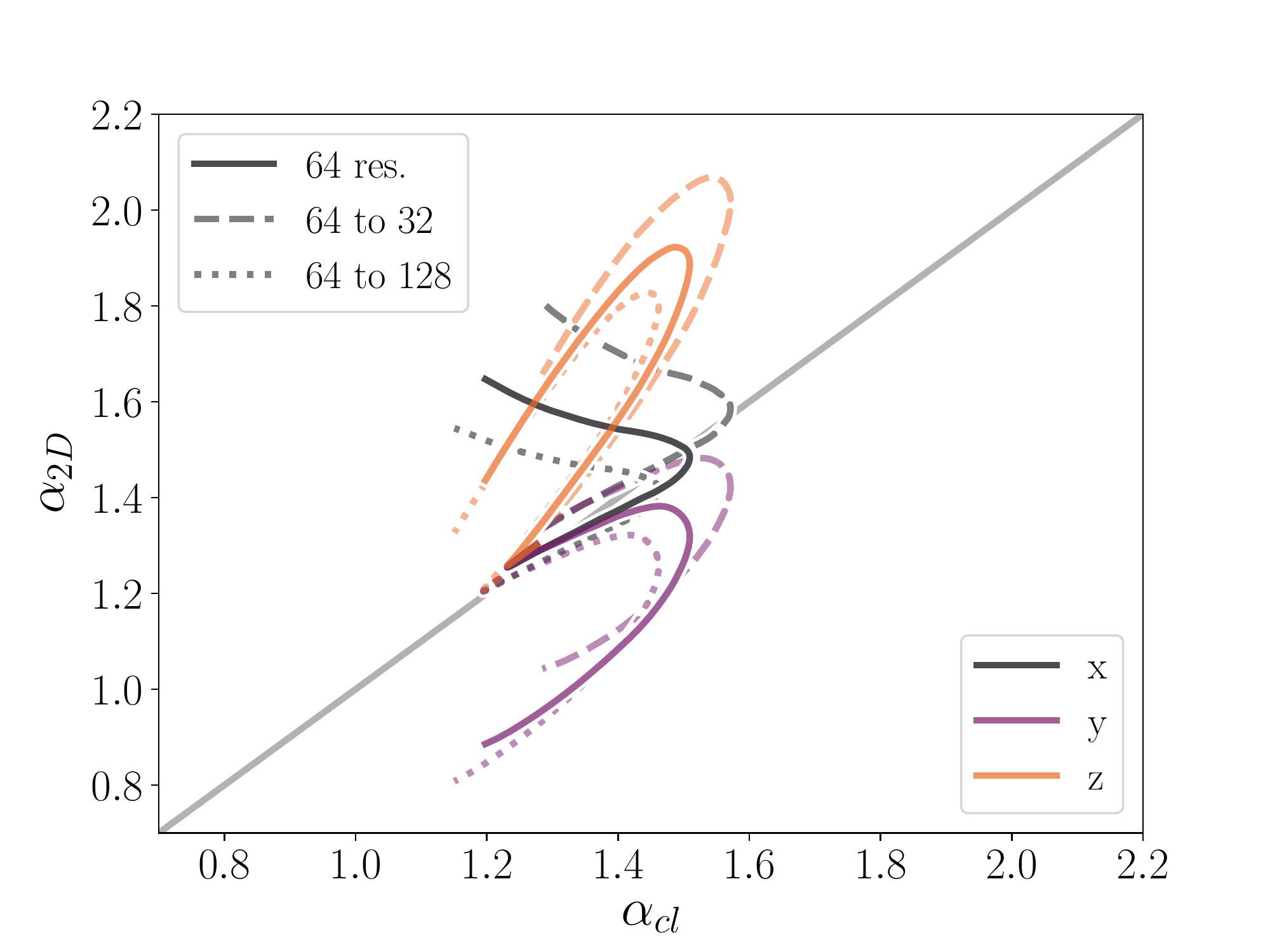} \\
\end{array}$
\caption{Values of $\alphatwod$ are compared with $\alphacl$ for the projections and resolutions displayed in Figure \ref{fig:energy_timescale}.  }
\label{fig:ratio_res}
\end{figure}

The first step is to assess how well $\calT$, $\calWcl$, and $\alphacl$ can be estimated from observations of a simulated cloud, under the idealization that the maps of the cloud column, velocity, and velocity dispersion are all known without any ambiguity.  The first two of these are inferred from three orthogonal projections of the same evolving cloud and plotted in Figure \ref{fig:energy_timescale}.  Focusing for the moment on the solid lines (native resolution results), we see that $\calTtwod$ and $\calWtwod$ deviate from $\calT$ and $\calWcl$, respectively, as the cloud collapses and becomes more filamentary.  The discrepancy is of order 20\% in each quantity over the course of this run.  
In Figure \ref{fig:ratio_res}, we compare $\alphacl$ with its observational estimate $\alphatwod$ (expression \ref{eq:VirialRatioEstimate}). {We note that while the two are clearly correlated, $\alphatwod$ exhibits errors of up to 40\%, comparable to the range of values spanned by $\alphacl$ within the simulation.}  

The run of $\alphacl$ in our simulation is quite limited compared to the range observed in molecular clouds, so we caution against drawing any strong conclusions from this exercise.  However, it provides a preview of what might be found in more wide-ranging simulations.

\subsection{Resolution effects}\label{SS:DifferentResolution}
The trends observed in Figures \ref{fig:energy_timescale} and \ref{fig:ratio_res} are due in part to the development of an elongated rotating structure during the cloud's collapse.  However, we require a means to separate this from finite-resolution effects, which also evolve during the simulation because our cloud occupies fewer resolution elements as it collapses.   Although we have performed runs at different numerical resolution, these cannot be directly compared  because they do not evolve in identical ways.  

Our approach, therefore, is to consider the same $64^3$-resolution run after artificially degrading each output to $32^3$ (by averaging), or enhancing it to $128^3$ (by cubic interpolation).  The results of this exercise are plotted as dashed and dotted lines, respectively.   We see that finite resolution most severely affects $\calWtwod/\calWcl$, and that a factor of four in linear resolution can have a $\sim 20\%$ effect on our inference of $\alpha$.   In fact our values for $\calWtwod$ and $\calWcl$ are both resolution dependent, as can be deduced from the diagonal shift with resolution in Figure \ref{fig:ratio_res}.  

These points are equally relevant to the observational determinations of $\calWtwod$ and $\alphatwod$, which must rely on maps of $\Sigmacl$ obtained at finite resolution.  

\subsection{Comparing $\alphatwod$ with $\alphaBM$}\label{SS:CompareAlphaBM}

\begin{figure}[ht]
\centering
$\begin{array}{c}
    \vspace{0.2cm}
    \hspace{-0.3cm}
    \includegraphics[scale=0.39]{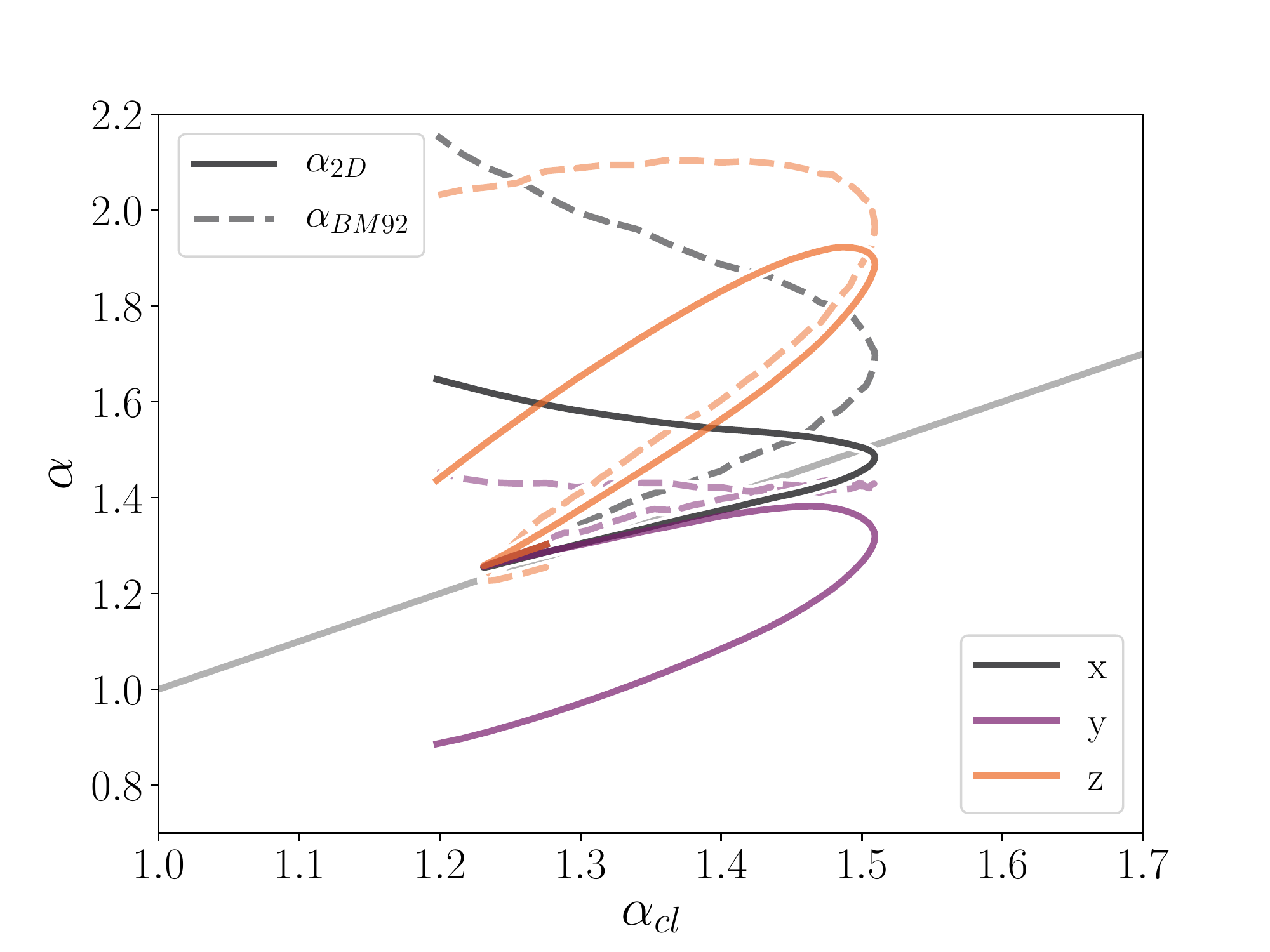} \\
\end{array}$
\caption{Values of $\alphatwod$ and $\alphaBM$ are compared with $\alphacl$ for the projections and resolutions displayed in Figure \ref{fig:energy_timescale}.  Note that the two quantities have distinct meanings, as discussed in \S\ref{SS:CompareAlphaBM}.}
\label{fig:diff_method}
\end{figure}

To ascertain whether it is worth the effort to compute $\alphatwod$, we provide a comparison with  $\alphaBM$ in Figure \ref{fig:diff_method}. 
To use equation (\ref{eq:alpha}) we require values for $\sigmaclz$, $\Rcl$ and  $\Mcl$.  For this comparison we adopt the same cloud boundary and cloud mass as for our computation of $\alphatwod$; we take
 $\sigmaclz$ to be the mean of $\sigmaclz(x,y)$ within the map; and we set $\Rcl$ so the cloud area is $\pi \Rcl^2$.
 
Figure \ref{fig:diff_method} shows that $\alphaBM$ and $\alphatwod$ both agree with $\alphacl$ in the initial state; this is expected, as a uniform spherical cloud has $a=1$.   Both quantities then deviate from $\alphacl$, with $\alphaBM$ increasing relative to $\alphatwod$ as the cloud's concentration increases and its surface becomes nonspherical.   

It is important to note that these quantities are not directly comparable, because $\alphatwod$ is a direct estimate for $\alphacl$, whereas $\alphaBM$ estimates $\alphacl/a$.   Figure \ref{fig:diff_method} shows, first, that both $\alphatwod$ and $\alphaBM$ are unavoidably affected by projection effects when only projected quantities are available; second, that $a$ increases as a cloud becomes more condensed; and third, that the value of $a$ itself depends on viewing angle.   We conclude $\alphatwod$ provides a measure of $a$, and that it is equally useful as a direct estimate of $\alphacl$.

\subsection{Calibration of techniques to disentangle $\Sigma$} \label{SS:ProjectionEffectsInSims}
An ability to estimate $2\calT/|\calWcl|$ from the projected cloud quantities is not much use if the cloud cannot be distinguished from other material along the line of sight.  Here, we evaluate the strategies to determine $\Sigmacl$ from $\Sigma$ discussed in \S \ref{S:DisentanglingSigma}. 

The first option (clipping) is to simply set $\Sigmacl=\Sigma$ within the cloud boundary, ignoring the problem.  The second (edge interpolation and subtraction) involves subtracting a smooth function that matches $\Sigma$ on the cloud boundary, such as a spline interpolation.   The third involves modeling the contribution of the envelope material and accounting for the fact that $\Sigma_{\rm env}$ should be greater at the cloud periphery than in the center; for this we use Abel's transformation as discussed in \S \ref{SS:Abel-theory}.

\begin{figure}[h]
\centering
$\begin{array}{c}
    \vspace{-0.2cm}
    \hspace{-0.15cm}
    \includegraphics[scale=0.45]{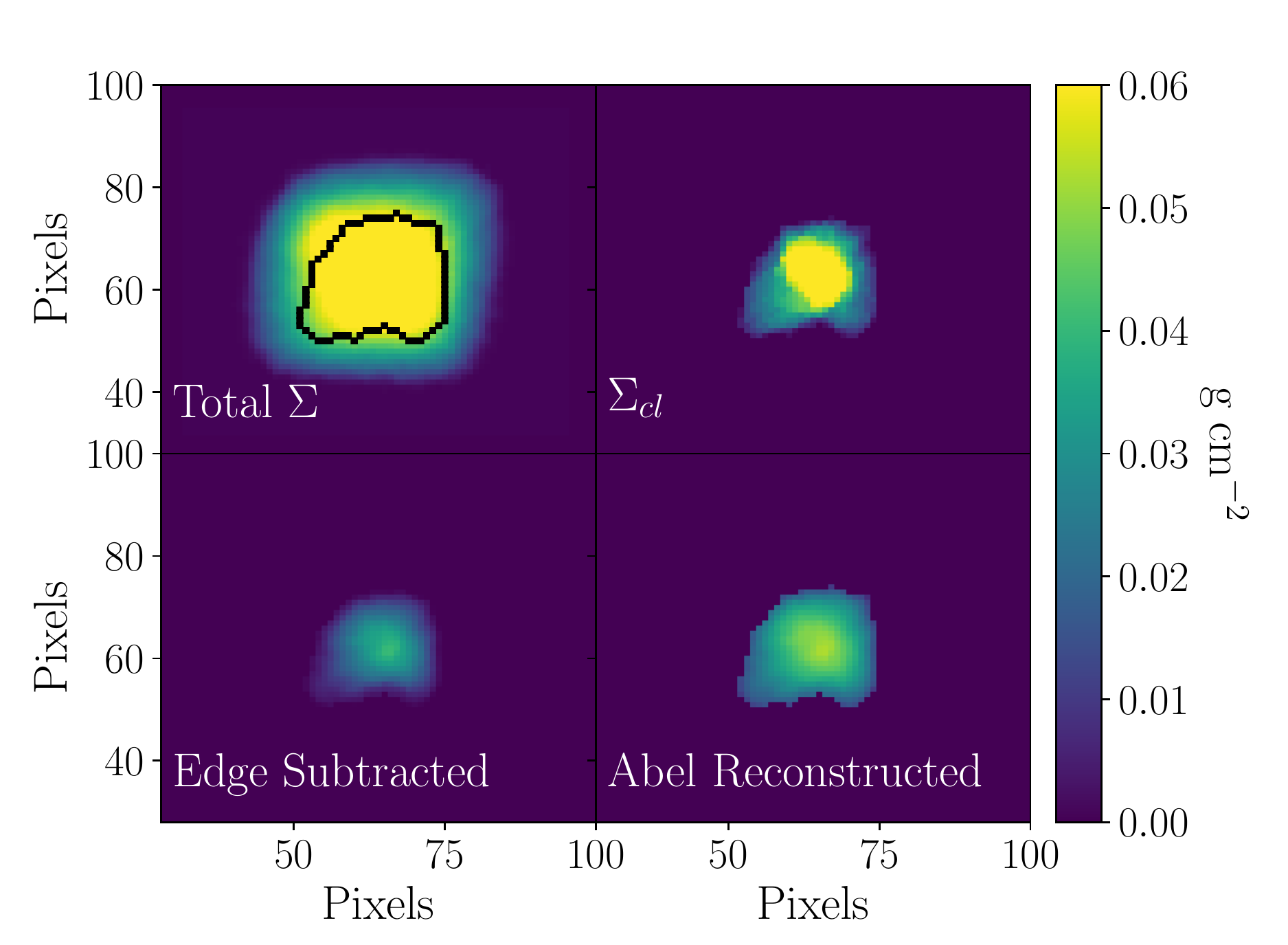} \\
\end{array}$
\caption{Maps of $\Sigma$ (top left) and $\Sigmacl$ (top right) along with reconstructions of $\Sigmacl$ described in \S \ref{S:DisentanglingSigma}.  }
\label{fig:cloudcutimage}
\end{figure}

\begin{figure}[h]
\centering
$\begin{array}{c}
    \vspace{-0.2cm}
    \hspace{-0.45cm}
    \includegraphics[scale=0.5]{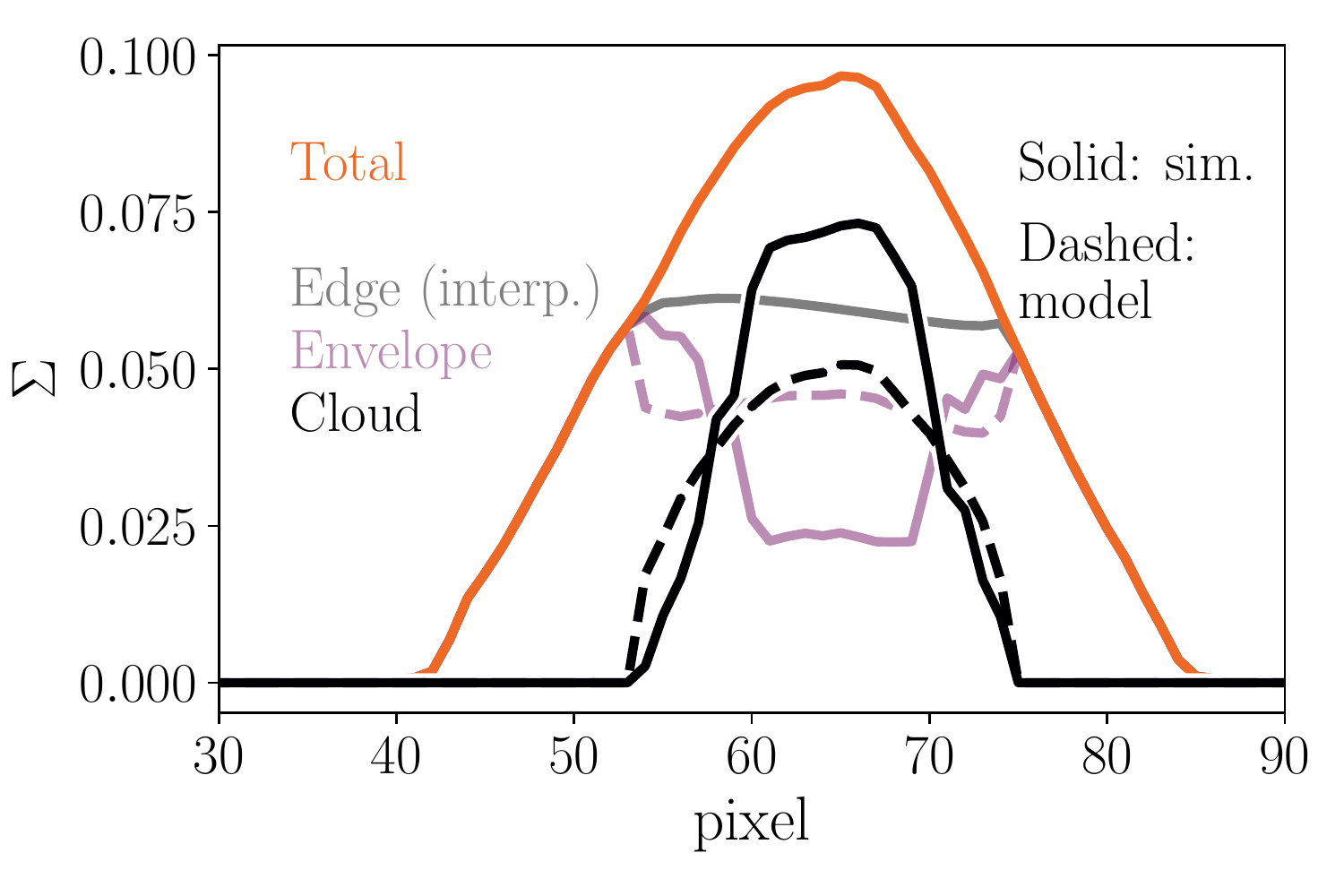} \\
\end{array}$
\caption{Profile of a slice through a middle of a map generated from our $64^3$ simulation at $t = 1.28~t_{\rm ff,0}$. The black line is $\Sigma_{cl}$; the grey line is our interpolated value of $\Sigma_{\rm edge}$;  $\Sigma_{\rm env}$ is shown in red and $\Sigmacl$ in purple.  Solid lines are computed directly from the simulation, while the dashed lines refer to reconstructions based on Abel's transformation.}
\label{fig:cloudcut}
\end{figure}

The top panel of Figure \ref{fig:cloudcutimage} displays $\Sigma$ and $\Sigmacl$ derived from a single time step of our simulation, along with the projected cloud boundary (black line in the top left panel).  Estimating $\Sigmacl$ by clipping is equivalent to considering only the region of $\Sigma$ interior to this boundary.  The bottom panels show two alternate methods of reconstruction: on the left, interpolation and subtraction of the edge value and on the right, reconstruction via Abel's transformation.  Figure \ref{fig:cloudcut} demonstrates a horizontal slice at the vertical midpoint, allowing for more quantitative comparisons.   As expected, clipping significantly over predicts $\Sigmacl$, while edge subtraction underestimates it.  The Abel reconstruction is also imperfect, in that it overestimates the envelope contribution at the cloud edge and underestimates it at the center, but it performs the best of these three methods.   This is apparent in Table \ref{table:energy-errors}, in which we compare the values of $\calTtwod$, $\calWtwod$, and $\alphatwod$ derived from these reconstructions with those from $\Sigmacl$. 

We note that the determination of $\tilde\rho(\tilde r)$  from $\tilde\Sigma(\tilde R)$ has the tendency to amplify noise, due to the derivative in Abel's transform (equation \ref{eq:abelbackward}).  This appears to be mitigated by projection, as we do not observe $\tilde\Sigmacl$ to be noisy.   However, the robustness of Abel reconstruction merits further study.

\begin{deluxetable*}{ccccc}
\tabletypesize{\footnotesize}
\tablecolumns{5} 
\tablewidth{0pt}
 \tablecaption{ Energy errors due to $\Sigmacl$ reconstruction
 \label{table:energy-errors}}
 \tablehead{
 \colhead{Quantity} & \colhead{Clipping } & \colhead{Edge subtraction } & \colhead{Abel reconstruction }& \colhead{True} 
 } 
\startdata 
  $\calTtwod/\calT$ &  1.9& 0.4&  0.8&0.8
  \\ 
$\calWtwod/\calWcl$ &  4.1&  0.2&  0.8&1.0
 \\ 
  $\alphatwod$ & 1.9 &   6.7 &  4.0 &3.2
\enddata
 \tablecomments{Error in inferred energies, and their virial ratio, arising from various techniques for disentangling $\Sigmacl$ from $\Sigma$, for the projection shown in Figure \ref{fig:cloudcutimage}.  Here `true' refers to $\Sigmacl$ derived from the three-dimensional simulation; all quantities are evaluated at the same, finite resolution. }
\end{deluxetable*}

\section{Discussion and conclusions}
\label{S:discuss}

We have addressed two sources of inaccuracy in the determination of the virial ratio from maps of an observed cloud: the estimation of the cloud's gravitational energy, and the separation of the cloud column from foreground and background matter.   For the former problem we propose an estimate based on the energy of a sheet with the same $\Sigmacl$.  For the latter, we propose a reconstruction based on Abel's transform.   Both techniques provide exact results for idealized cases (perfectly resolved spherical and axisymmetric clouds, respectively) and appear to perform well in general.  However, the effects of projection lead to significant errors for general three-dimensional structures, and we suspect these will be difficult to eliminate without additional information about the line-of-sight distribution of matter.  

We envision two possible sources for this additional information, which should be the topic of future work. First, it may be that \textit{Gaia} or a subsequent instrument will provide enough distances and extinctions to sources inside and surrounding a cloud for the three-dimensional distribution of dust to be mapped.  This would also solve the problem of foreground and background subtraction and relieve distance ambiguities.  Second, we have so far ignored the fact that molecular line emission also constrains the mass distribution (albeit in ways that depend on chemistry, excitation, and radiative transfer). 

We find that finite resolution can lead to an underestimate of the gravitational binding energy, an effect that should be considered (and perhaps, statistically removed) when these techniques are applied to observations. We have not considered several additional difficulties that lie outside the scope of this paper, such as the contamination of cloud velocity and velocity dispersion maps by emission from foreground and background material, or the effects of excitation and radiation transfer on the tracer. In upcoming papers, we plan to further calibrate the  techniques proposed here, and to apply them to molecular cloud data. 

\acknowledgements We thank Chris McKee and the anonymous referee for their comments, and our colleagues in the Green Bank Ammonia Survey project for their encouragement and feedback. This work was supported by an NSERC Discovery Grant (CDM).  Finally, we acknowledge that this work was conducted on the traditional land of the Huron-Wendat, the Seneca, and most recently, the Mississaugas of the Credit River; we are grateful for the opportunity to work here. 

\software{Flash \citep{Fryxell2000}}

\begin{appendix}
We prove relation (\ref{eq:2d-to-3d-ratio}) between $\calWtwod$ and $\calWcl$ more rigorously than the argument in \S~\ref{S:GravEstimation}.   The former can be written as a double integral, 
\begin{equation}\label{eq:Wcl2d_doubleintegral}
\calWtwod = -\frac2\pi\times \frac{G}{2}\iint_{A_{\rm cl}} {\Sigma(\wvec_1) \Sigma(\wvec_2)\over |\wvec_1-\wvec_2|} d^2\wvec_1 d^2\wvec_2
\end{equation}
where $\wvec = (x,y)$ is the projection of $\rvec=(x,y,z)$ in the plane of the sky.  Expanding each instance of $\Sigma(\wvec)$ according to its definition as $\int_{V_{\rm cl}} \rho(x,y,z)\, dz$, we recognize that the integrals $d^2\wvec\, dz=d^3\rvec$ are equivalent to integrals over the cloud volume: 
\begin{equation}\label{eq:Wcl2d_doubleintegral_B}
\calWtwod = -\frac2\pi\times\frac{G}{2}\iint_{V_{\rm cl}}d^3\rvec_1  d^2\rvec_2 {\rho(\rvec_1) \rho(\rvec_2)\over |\wvec_1-\wvec_2|}.
\end{equation}
In the denominator, $|\wvec_1-\wvec_2| = |\rvec_1-\rvec_2| (1-\mu_{12}^2)^{1/2}$, where $\mu_{12}$ is the angle cosine between $\rvec_1-\rvec_2$ and $\hat z$.  Therefore, 
\begin{eqnarray}\label{eq:Wcl2d_doubleintegral_C}
\langle\calWtwod\rangle &=& -\frac2\pi\times\frac{G}{2}\langle \iint_{V_{\rm cl}}d^3\rvec_1 d^2\rvec_2 {\rho(\rvec_1) \rho(\rvec_2)\over |\rvec_1-\rvec_2| (1-\mu_{12}^2)^{1/2} } \rangle  \\ 
&=& -\frac2\pi\times\frac{G}{2}  \iint_{V_{\rm cl}}d^3\rvec_1 d^2\rvec_2 {\rho(\rvec_1) \rho(\rvec_2)\over |\rvec_1-\rvec_2| } \langle (1-\mu_{12}^2)^{-1/2} \rangle; \nonumber
\end{eqnarray}
where the second line follows from the fact that the orientation average acts only on the term $(1-\mu_{12}^2)^{-1/2}$.   Furthermore $\mu_{1,2}$ is uniformly distributed from -1 to 1 regardless of the direction $\rvec_1-\rvec_2$, so the result of this average is $\pi/2$ in all cases; this cancels the prefactor $2/\pi$. The remainder of the integral is equivalent to $\calWcl$, so we arrive at the result in equation (\ref{eq:2d-to-3d-ratio}). 
\end{appendix}

\bibliographystyle{apj} 
\bibliography{Virial_Parameter}

\end{document}